\documentclass{article}

\begin{document}

\title{Parametric Level Correlations in Random--Matrix Models}

\author{Hans A. Weidenm\"uller \\
Max--Planck--Institut f\"ur Kernphysik, Heidelberg, Germany}

\maketitle

\ \ \ {\it Dedicated to Lothar Sch\"afer on the occasion of his
sixtieth birthday}

\abstract{We show that parametric level correlations in random--matrix
theories are closely related to a breaking of the symmetry between the
advanced and the retarded Green's functions. The form of the
parametric level correlation function is the same as for the
disordered case considered earlier by Simons and Altshuler and is
given by the graded trace of the commutator of the saddle--point
solution with the particular matrix that describes the symmetry
breaking in the actual case of interest. The strength factor differs
from the case of disorder. It is determined solely by the Goldstone
mode. It is essentially given by the number of levels that are
strongly mixed as the external parameter changes. The factor can
easily be estimated in applications.}

\section{Introduction}
\label{int}

Parametric level correlations in chaotic and disordered systems have
received much attention in the early 1990's (see the review~\cite{GMW}
and references therein). This development culminated in the seminal
work of Simons and Altshuler~\cite{SIM} who showed that such
correlations have a universal form and who calculated some of the
correlation functions for disordered systems explicitly.

In the present paper, I take a fresh look at this problem. This is
motivated by two circumstances. (i) The work of Simons  and Altshuler 
does not address level correlations of random matrices (but rather of
chaotic and/or disordered systems). But level correlations of random
matrices do play a role in some applications of random--matrix theory.
A case in point concerns correlations of levels with different spins
in atomic and nuclear systems~\cite{PAP}. While the {\it form} of the
correlation function obtained by Simons and Altshuler is expected to
be unchanged, it is necessary to determine the dimensionless parameter
which governs its behavior, and to connect that parameter with physical
parameters of the system at hand. In particular, the concept of ``level
velocities'' introduced by Simons and Altshuler needs to be
reconsidered. It will be shown that in contrast to the case of
disorder, the strength parameter in random--matrix theory is not
influenced by a coupling of the Goldstone mode with the massive modes.
(ii) Parametric level correlations can be seen as a manifestation of
symmetry breaking. The broken symmetry is that between the advanced
and the retarded Green's functions. I aim at a presentation which
displays this fact as clearly as possible. With this insight, writing
down the form of the correlation funtion is quite straightforward.

The correlation functions will be given for the GOE and for the GUE.
The results are also compared with the two--point correlation function
for the GOE $\to$ GUE transition caused by time--reversal symmetry
breaking. We shall see that in the latter case, symmetry breaking acts
differently.

\section{Formulation of the Problem}
\label{pro}

The ensemble of Hamiltonians $H$ has the form
\begin{equation}
H = H_1 \cos(X) + H_2 \sin(X)
\label{1a}
\end{equation}
where $X$ is a dimensionless parameter and where $H_1$ and $H_2$ are
uncorrelated random matrices belonging to the same symmetry class
of one of Dyson's three canonical ensembles. We wish to calculate the
parametric correlation function
\begin{equation}
k = \overline{ {\rm tr} [ \frac{1}{E^+_1 - H(X)} ] {\rm tr} [ \frac{1}
{E^-_2 - H(X')} ] } \ .
\label{2a}
\end{equation}
The overbar denotes the ensemble average. The function $k$ contains
quantitative information about the way in which the spectra at
parameter values $X$ and $X'$ are correlated. For $X = X'$, $k$
coincides with the standard two--point correlation function.

Both in the case of disordered systems and in the present case, one
needs to calculate $k$ only for small values of $|X - X'|$, i.e.,
perturbatively. The reason is that we are interested in local (rather
than global) changes of the spectrum. The former involve an energy
scale of order $d$, the mean level spacing, the latter, an energy scale
of order $N d$ where $N \to \infty$ is the dimension of the matrices
$H_1$ and $H_2$. Then, the function $k = k(\epsilon, X - X')$ depends
only upon the difference $\epsilon = E_1 - E_2$ of the energies of the
two Green's functions.

I expand the Hamiltonians $H(X)$ and $H(X')$ in Eq.~(\ref{1a}) around
the mid--point $X_0 = (1/2)(X + X')$ in powers of $X - X_0 = (1/2)(X
- X')$ and of $X' - X_0 = (1/2)(X' - X)$, respectively, and keep only
terms up to first order in $(X - X')$. Then,
\begin{eqnarray}
H(X) &\approx& H_0 + (1/2) (X - X') V \ , \nonumber \\
H(X') &\approx& H_0 - (1/2) (X - X') V \ ,
\label{3a}
\end{eqnarray}
where $H_0 = H(X_0)$ and where
\begin{equation}
V = H_2 \cos(X_0) - H_1 \sin(X_0) \ . 
\label{4a}
\end{equation}
The random matrices $H(X_0)$ and $V$ are uncorrelated,
\begin{equation}
\overline{H_0 V} = 0 \ .
\label{5a}
\end{equation}
This follows from the fact that $H_1$ and $H_2$ are uncorrelated,
$\overline{H_1 H_2} = 0$.

To identify the small parameter of the expansion, I define (as usual)
the spreading width due to the perturbation as
\begin{equation}
\Gamma^\downarrow = 2 \pi (X - X')^2 \overline{V^2} / d \ .
\label{6a}
\end{equation}
The spreading width is a measure of the energy interval within which
the levels of $H_0$ get strongly mixed as the parameter changes from
$X$ to $X'$. We are interested in values of $\Gamma^\downarrow$ which
are of the order of $d$ (rather than $N d$). We normalize the variances
of $H_1$ and $H_2$ in the usual manner,
\begin{equation}
\overline{{(H_j)}_{\mu \nu} {(H_j)}_{\nu \mu}} = \frac{\lambda^2}{N}
\ ; \ j = 1,2 \ ; \mu \neq \nu \ ,
\label{7a}
\end{equation}
where $\mu$ and $\nu$ are level indices, and where $2 \lambda$ is the
radius of the semicircle. The mean level spacing of $H_0$ in the centre
of the semicircle is given by $d = \pi \lambda / N$. With these
conventions, we have
\begin{equation}
\Gamma^\downarrow = 2 (X - X')^2 \lambda \ .
\label{8a}
\end{equation}
We shall see that the dimensionles parameter which governs the level
correlation function is given by $\Gamma^\downarrow / d = (2/\pi) N
(X - X')^2$. For this parameter to be of order unity, we must have
that $(X - X')^2$ is of order $1/N$. This jusitifies our perturbation
expansion and the fact that we keep only linear terms in $(X - X')$. 

Substituting $H(X)$ and $H(X')$ from Eqs.~(\ref{3a}) into Eq.~(\ref{2a})
yields
\begin{eqnarray}
&& k(\epsilon, X - X') \nonumber \\
&& = \overline{ {\rm tr} [ \frac{1}{E^+_1 - H_0 - (1/2)(X - X') V} ]
{\rm tr} [ \frac{1}{E^-_2 - H_0 + (1/2)(X - X') V} ] } \ . \nonumber \\
\label{9a}
\end{eqnarray}
Eq.~(\ref{9a}) displays explicitly the fact that the perturbation $V$
breaks the symmetry between the retarded and the advanced Green's
functions. This is essential for the supersymmetry calculation of
$k(\epsilon, X - X')$.

\section{Supersymmetry}
\label{sup}

The supersymmetry method~\cite{efe,vwz} has become a standard tool in
random--matrix theory. Therefore, I confine myself to giving the
essential steps in the calculation. I do so for the case where both
$H_1$ and $H_2$ belong to the GOE, and give only results for the GUE.

I proceed as in Ref.~\cite{vwz}, also use their notation, and arrive
at the following form of the generating function,
\begin{equation}
Z(E_1, E_2; X, X', J) = \int {\rm d} [\Psi] \exp \{ {\cal L} (\Psi, J)
\} \ ,
\label{3}
\end{equation}
where the Lagrangian is given by
\begin{equation}
{\cal L} = (1/2) i ( \Psi^\dagger L^{1/2} D^J L^{1/2} \Psi ) \ .
\label{4}
\end{equation}
Here $D^J$ is a graded matrix of dimension 8, given by
\begin{equation}
D^J = ({\bf E} - {\bf H} + i \delta + {\bf J} - (1/2) {\cal E}) \ .
\label{5}
\end{equation}
According to Eq.~(\ref{3a}), the matrix ${\bf H}$ has the form
\begin{equation}
{\bf H} =  H(X_0) {\bf 1}_8 + (1/2) (X - X') V L \ .
\label{6}
\end{equation}
Here ${\bf 1}_8$ denotes the unit matrix in eight dimensions, while
\begin{equation}
L = {\rm diag}(1, 1, 1, 1, -1, -1, -1, -1)
\label{6b}
\end{equation}
is the matrix which breaks the symmetry between the advanced and the
retarded Green's functions.
We want to calculate the two--point function and accordingly put
\begin{equation}
{\bf J} = \delta_{\mu \nu} {\rm diag} (-j_1, -j_1, +j_1, +j_1, -j_2,
-j_2, +j_2, +j_2) = \delta_{\mu \nu} ({\bf j}_1, {\bf j}_2) \ . 
\label{6c}
\end{equation}
The last equation defines $({\bf j}_1, {\bf j}_2)$.

The ensemble average is given in terms of the second moment of the
term $(i/2) (\Psi^\dagger L^{1/2} {\bf H} L^{1/2} \Psi )$,
\begin{eqnarray}
&& \overline{ [ (i/2) (\Psi^\dagger L^{1/2} {\bf H} L^{1/2} \Psi ) ]^2
  } \nonumber \\
&&= - (\lambda^2/(2N)) \sum_{\mu \nu \alpha \beta} \biggl( \Psi
  ^\dagger_{\mu \alpha} (L^{1/2})_{\alpha \alpha} (L^{1/2})_{\alpha
  \alpha} \Psi_{\nu \alpha} \biggr) \nonumber \\
&& \qquad \qquad \times \biggl( \Psi^\dagger_{\nu \beta} (L^{1/2})
  _{\beta \beta} (L^{1/2})_{\beta \beta} \Psi_{\mu \beta} \biggr)
  \nonumber \\
&& \qquad - (\lambda^2/(8N)) (X - X')^2 \sum_{\mu \nu \alpha \beta}
  \biggl( \Psi^\dagger_{\mu \alpha} (L^{1/2})_{\alpha \alpha} L_{\alpha
  \alpha} (L^{1/2})_{\alpha \alpha} \Psi_{\nu \alpha} \biggr)
  \nonumber \\
&& \qquad \qquad \times \biggl( \Psi^\dagger_{\nu \beta} (L^{1/2})
  {\beta \beta} L_{\beta \beta} (L^{1/2})_{\beta \beta} \Psi_{\mu \beta}
  \biggr) \ .
\label{28}
\end{eqnarray}
The summation over $\alpha, \beta$ runs from 1 to 8, that over $\mu,
\nu$ from 1 to $N$. I define
\begin{eqnarray}
A_{\alpha \beta} &=& i \lambda \sum_{\mu} (L^{1/2})_{\alpha \alpha}
\psi_{\mu \alpha} \psi^\dagger_{\mu \beta} (L^{1/2})_{\beta \beta}
\nonumber \\
&& + (1/8) (X - X')^2 i \lambda \sum_{\mu} L_{\alpha \alpha}
(L^{1/2})_{\alpha \alpha} \psi_{\mu \alpha} \psi^\dagger_{\mu \beta}
(L^{1/2})_{\beta \beta} L_{\beta \beta} \ . \nonumber \\
\label{29}
\end{eqnarray}
This equation clearly displays the separate contributions from $H_0$
and from the symmetry--breaking term $V L$. Under neglect of
higher--order terms in $(X - X')^2$ (which we have shown to be
negligible for $N \to \infty$), the right--hand side of Eq.~(\ref{28})
can be expressed in terms of $A$, yielding
\begin{equation}
\overline{ [ (i/2) (\Psi^\dagger L^{1/2} {\bf H} L^{1/2} \Psi ) ]^2 }
= \frac{1}{2N} {\rm trg}_\alpha (A^2) \ .  
\label{30}
\end{equation}
The Hubbard--Stratonovitch transformation yields now for $Z$ the form
\begin{equation}
Z(E_1, E_2; X, X', J) = \int {\rm d} [\sigma] \exp \biggl \{ -
\frac{N}{4} {\rm trg}_{\alpha} (\sigma^2) - \frac{N}{2} {\rm
trg}_{\alpha} \ln {\bf N}(J) \biggr \} \ ,
\label{9}
\end{equation}
where
\begin{equation}
{\bf N}(J) = E {\bf 1}_8 - (1/2) {\cal E} + i \delta - \lambda \Sigma +
({\bf j}_1,{\bf j}_2)
\label{10}
\end{equation}
and
\begin{equation}
\Sigma = \sigma + (1/8) (X - X')^2 \ L \sigma L \ .  
\label{31}
\end{equation}
We use the saddle--point approximation, omitting terms which are of
order $1/N$. These are the terms proportional to ${\cal E}$, to
$(X' - X)^2$, and the source terms. The saddle--point equation
\begin{equation}
\sigma = \frac{\lambda}{E {\bf 1}_8 - \lambda \sigma}
\label{17}
\end{equation}
has the standard solution
\begin{equation}
\sigma_G = T^{-1}_0 \sigma^0_D T_0
\label{18}
\end{equation}
with $\sigma^0_D$ diagonal and given by
\begin{equation}
\sigma^0_D = \frac{E}{2 \lambda} - i \Delta_0 L
\label{19}
\end{equation}
and $\Delta_0 = \sqrt{1 - (E/(2 \lambda))^2}$. The full sigma matrix
is written as
\begin{equation}
\sigma = \sigma_G + \delta \sigma = \sigma_G +  T^{-1}_0 \delta P T_0
\ .
\label{20}
\end{equation}
It remains to work out the integrals over the massive modes, and over
the Goldstone mode.

\section{Integration over the Massive Modes}
\label{mas}

In applications of the supersymmetry formalism, one would normally
skip the present Section because the integration over the massive modes
is known to simply yield a constant. However, in the work of Simons and
Altshuler~\cite{SIM}, it is shown that the strength of the parametric
level correlation function depends upon contributions due to the
coupling of the Goldstone mode with the massive modes. Is such a
mechanism also operative in the present case? To answer this question,
I substitute in Eqs.~(\ref{9}) and (\ref{10}) for $\Sigma$ the
expression~(\ref{31}) and in the latter for $\sigma$ the
expression~(\ref{20}). I expand in powers of $\delta \sigma$ and of the
small entities ${\cal E}, (X' - X)^2$ and $({\bf j}_1,{\bf j}_2)$ and
keep terms up to the second order in $\delta \sigma$ and up to first
order in the other small entities. Some of the linear terms in $\delta
\sigma$ cancel because of the saddle--point condition. The exponent in
Eq.~(\ref{10}) takes the form
\begin{eqnarray}
&& - \frac{N}{4} {\rm trg}_\alpha \biggl[ \delta \sigma \biggr]^2 +
  \frac{N}{4} {\rm trg}_\alpha \biggl[ \sigma_G \delta \sigma \biggr]^2
  + \frac{N \epsilon}{4 \lambda} {\rm trg}_\alpha \biggl[ \sigma_G L
  \biggr] \nonumber \\
&& \qquad - \frac{N}{2 \lambda} {\rm trg}_\alpha \biggl[ \sigma_G ({\bf
  j}_1, {\bf j}_2) \biggr] + \frac{N}{16} (X - X')^2 {\rm trg}_\alpha
  \biggl[ (\sigma_G L)^2 \biggr] \nonumber \\
&& \qquad + \frac{N}{16} (X - X')^2 {\rm trg}_\alpha \biggl[ \sigma_G
  L \delta \sigma L \biggr] + \frac{N \epsilon}{4 \lambda} {\rm trg}
  _\alpha \biggl[ \sigma_G \delta \sigma \sigma_G L \biggr] \nonumber \\
&& \qquad - \frac{N}{2 \lambda} {\rm trg}_\alpha \biggl[ \sigma_G \delta
  \sigma \sigma_G ({\bf j}_1, {\bf j}_2) \biggr] \nonumber
  \\
&& \qquad + \frac{N}{16} (X - X')^2 {\rm trg}_\alpha \biggl[ \sigma_G
  \delta \sigma (\sigma_G L)^2 + \sigma_G \delta \sigma \sigma_G L \delta
  \sigma L \biggr] \nonumber \\
&& \qquad + \frac{N \epsilon}{4 \lambda} {\rm trg}_\alpha \biggl[
  \sigma_G L (\sigma_G \delta \sigma )^2 \biggr] - \frac{N \epsilon}{2
  \lambda} {\rm trg}_\alpha \biggl[ \sigma_G ({\bf j}_1, {\bf j}_2)
  (\sigma_G \delta \sigma )^2 \biggr] \nonumber \\
&& \qquad + \frac{N}{16} (X - X')^2 {\rm trg}_\alpha \biggl[ ( \sigma_G
  L )^2 (\sigma_G \delta \sigma )^2 \biggr] \ .    
\label{21}
\end{eqnarray}
The leading terms in $(\delta \sigma)^2$ are the first two terms in
expression~(\ref{21}). These terms show that all massive modes have
mass $N$. We recall that $N ( X - X')^2$ is of order unity. Therefore,
the remaining terms which are quadratic in $\delta \sigma$ are
negligible. The terms linear in $\delta \sigma$ are all at most of
order unity. To be non--negligible, they ought to be of order
$\sqrt{N}$. In the limit $N \to \infty$ we are, thus, left with the
first five terms in expression~(\ref{21}). This shows that the massive
modes decouple from the Goldstone mode. Moreover, the massive--mode
contribution attains exactly the form given in Ref.~\cite{vwz} and
can, therefore, be integrated out without any problem. Hence, in
contrast to the disorder problem studied in Ref.~\cite{SIM}, the
massive modes do not contribute to the strength of the parametric
level correlation function in random--matrix theory.

The result for $Z$ is
\begin{eqnarray}
Z(E_1, E_2; X, X', J) &=& 4 \int {\rm d} [\sigma] \exp \biggl \{ +
\frac{\pi \epsilon}{4 d} {\rm trg}_{\alpha} ( \sigma_G L ) \nonumber \\
&& + \frac{N}{16} (X - X')^2 {\rm trg}_\alpha ( \sigma_G L )^2
\biggr \} \nonumber \\
&& \times \frac{N^2}{8 \lambda^2} \biggl ( {\rm trg}_{\alpha} [ ({\bf
  j}_1, {\bf j}_2 ) \sigma_G ] \biggr )^2 \ .
\label{22}
\end{eqnarray}
I carry out the differentiation with respect to $j_1$ and $j_2$. The
result is
\begin{eqnarray}
k(\epsilon, X - X') &=& (1/2) \int {\rm d} [\sigma] \exp \biggl \{ +
\frac{\pi \epsilon}{4 d} {\rm trg}_{\alpha} ( \sigma_G L ) \nonumber \\
&& + \frac{N}{16} (X - X')^2 {\rm trg}_\alpha ( \sigma_G L )^2 \biggr
\} \nonumber \\  
&& \times \frac{\pi^2}{d^2} ( {\rm trg}_{\alpha} [ I(1) (\sigma_G)_{1,
  1} ] ) ( {\rm trg}_{\alpha} [ I(2) (\sigma_G)_{2, 2} ] ) \ .
\label{23}
\end{eqnarray}

\section{Integration over the Goldstone Mode}
\label{gol}

Our result Eq.~(\ref{23}) differs from the standard expression for the
GOE two--point function by an additional term appearing in the exponent.
Using Eq.~(\ref{8a}), we rewrite this term in the form
\begin{equation}
+ \frac{N}{16} (X - X')^2 {\rm trg}_\alpha ( \sigma_G L )^2 =
+ \frac{\pi \Gamma^\downarrow}{64 d} {\rm trg}_{\alpha} [ ( [ \sigma_G,
 L ] )^2 ] \ .
\label{25}
\end{equation}
Once again, the right--hand side of this equation shows very clearly
that the term is due to the symmetry breaking caused by the
perturbation.

In the three graded traces appearing on the right--hand side of
Eq.~(\ref{23}), the only matrices which break the pseudounitary
symmetry are $I(1)$ and $I(2)$. Therefore, the two graded traces in
the exponent depend only upon the ``eigenvalues'' (remaining
integration variables). I use the parametrizations of both
Refs.~\cite{vwz} and \cite{efe} to work out $Z$ in the middle of
the spectrum where $\Delta_0 = 1$.

For the parametrization of Ref.~\cite{vwz} I find
\begin{eqnarray}
k(\epsilon, X - X') &\propto& \int_0^\infty {\rm d} \lambda_1
\int_0^\infty {\rm d} \lambda_2 \int_0^1 {\rm d} \lambda \nonumber \\
&& \times \frac{(1 - \lambda) \lambda |\lambda_1 - \lambda_2|}{((1 +
  \lambda_1) \lambda_1 (1 + \lambda_2) \lambda_2)^{1/2} (\lambda +
  \lambda_1)^2 (\lambda + \lambda_2)^2} \nonumber \\
&& \times \exp \biggl \{ - \frac{i \pi \epsilon}{d} ( \lambda_1 +
\lambda_2 + 2 \lambda ) \nonumber \\
&& \qquad - \frac{\pi \Gamma^\downarrow}{4 d} (\lambda_1 + \lambda_2 +
2 \lambda ) ( 1 + \lambda_1 + \lambda_2 + 2 \lambda ) \biggr \}
\nonumber \\
&& \times ( \lambda_1 + \lambda_2 + 2 \lambda )^2 \ .
\label{38a}
\end{eqnarray}
For the parametrization of Ref.~\cite{efe}, the two terms in the
exponent take the form
\begin{equation}
\frac{i \pi \omega}{d} (\lambda - \lambda_1 \lambda_2)
- \frac{\pi \Gamma^\downarrow}{4 d} (2 \lambda_1^2 \lambda_2^2 -
  \lambda_1^2 - \lambda_2^2  - \lambda^2 + 1 ) \ .
\end{equation}
The first term agrees with Efetov's Eq.~(5.35) if the definition of
$x$ following this equation is taken into account. The combination of
integration variables appearing in the second term is the same as
given by Simons and Altshuler.

Hopefully, the derivation given above shows very clearly the role of
symmetry breaking in parametric level correlations. The result
confirms our expectation: The form of the parametric level correlation
function is the same as for the disordered case. The strength factor
differs and is given by $\pi \Gamma^{\downarrow} / (4 d)$. Except for
the numerical factor $\pi / 4$, this result, too, corresponds to naive
expectations: $\Gamma^{\downarrow} / d$ is a measure of the number of
levels which are strongly mixed with each other as the external
parameter changes from $X$ to $X'$. In applications, this paramter can
be estimated in terms of the strength of the perturbation and of the
local mean level spacing.

\section{General Aspects of Symmetry Breaking}
\label{sym1}

I now address more fully the symmetry--breaking mechanism which
occurs when one considers parametric level correlations between two
Hamiltonian ensembles $H_1$ and $H_2$ (symbolically denoted by $H_1
\longleftrightarrow H_2$). I do so in several situations with the
intention of exhibiting the underlying similarities and differences.
I consider the following cases: (i) GOE $\longleftrightarrow$ GOE;
(ii) GUE $\longleftrightarrow$ GUE; (iii) GOE $\longleftrightarrow$
GUE. In the last case, the two Hamiltonians $H_1$ and $H_2$ obviously
do not belong to the same symmetry class. For the sake of comparison,
I consider also (iv) the two--point autocorrelation function for the
GOE $\to$ GUE transition~\cite{alt}. Proceeding as before, I calculate
the resulting contributions to the effective Lagrangean (i.e., the
additional symmetry--breaking terms in the exponent). These are
jointly denoted by $S$ and referred to as the parametric correlator.
The commuting (c) and anticommuting (a) integration variables are
arranged as follows: For the GUE two--point function, the sequence is
(c, a, c, a) while for the GOE, it is (c, c, a, a, c, c, a, a). I
define the graded matrix
\begin{equation}
{\rm T}_3 = {\rm diag} (+1, -1, +1, -1; +1, -1, +1, -1) \ .
\label{44}
\end{equation}
Case (i) has been considered above. The parametric correlator was
found to have the form $S_{({\rm i})} = {\rm trg} \{ ( [\sigma_G, L]
  )^2 \}$. Case (ii) is formally very similar and leads to the same
expression except that now $\sigma_G$ and $L$ have dimension four
rather than eight. In case (iii), we get the GUE by adding to the GOE
matrix in the advanced Green's function an imaginary random matrix.
Thus, the new term in the exponent arises only from the advanced
Green's function (and not from both, the retarded and the advanced
Green's function as in the previous cases (i) and (ii)). Moreover,
the new term carries the matrix $\tau_3$ as the signal for the
breaking of GOE symmetry and suppression of the Cooperon mode. As a
result, the relevant term has the form $S_{({\rm iii})} = {\rm trg}
\{ ( [\sigma_G, ({\bf 1}_8 - L_8) {\rm T}_3 ])^2 \}$. Case (iv) leads
to a symmetry--breaking term of the form $S_{({\rm iv})} = {\rm trg}
\{ ( [\sigma_G, {\rm T}_3] )^2 \}$. This obviously differs from
$S_{({\rm iii})}$. I observe that all these paramertic correlators
have the same form, ${\rm trg} \{ ( [ \sigma_G, T_{\rm x} ] )^2 \}$,
with $T_{\rm x}$ given by
\begin{eqnarray}
T_{\rm i} &=&  L_8 \ , \nonumber \\
T_{\rm ii} &=& L_4\ , \nonumber \\
T_{\rm iii} &=& (1_8 - L_8) {\rm T}_3 \ , \nonumber \\ 
T_{\rm iv} &=& {\rm T}_3 \ .
\label{45}
\end{eqnarray}

In summary, we have shown that the parametric correlation functions in
random--matrix theory have a very simple form. Each one is obtained
from the standard two--point function for level correlations by adding
in the exponent of the generating function a term. That term is given
by the graded trace of the commutator of the saddle--point solution
$\sigma_G$ with the particular matrix that describes the symmetry
breaking in the actual case of interest. Except for a numerical factor
which is of order unity, the factor in front of the commutator is
given by $\Gamma^{\downarrow} / d$.

{\bf Acknowledgment}. I am grateful to T. Papenbrock and Z. Pluhar for
helpful discussions, and to T. Papenbrock for valuable suggestions.

\end{document}